\documentclass[12pt,reqno]{article}
\usepackage{amsmath, amsthm, amssymb,epsfig,alltt}

\def\a{\alpha}
\def\b{\beta}

\def\k{\kappa}

\def\e{\epsilon}
\def\p{\partial}
\def\m{\mu}
\def\n{\nu}
\def\t{\tau}
\def\th{\theta}
\def\s{\sigma}
\def\g{\gamma}

\def\half{\frac{1}{2}}

\def\sp{\sigma^\prime}
\def\nn{\nonumber}

\def\barxi{{\bar \xi}}
\def\2pap{2\pi\alpha^\prime}

\def\beq{\begin{eqnarray}}
 \def\eeq{\end{eqnarray}}
 \def\4pap{4\pi\a^\prime}

 \def\sp{{\s^\prime}}
 \def\spp{\s^{\prime\prime}}
 \def\ap{{\a^\prime}}

 \def\barpsi{{\bar \psi}}

 \def\bbZ{\mathbb{Z}}
 \def\bolG{\boldsymbol G}
\begin{document}


\title{$U(1)$ Chiral Symmetry in  
One-Dimensional \\ Interacting Electron System with Spin}

\author{Taejin Lee
\\~\\
Department of Physics, Kangwon National University \\
Chuncheon 200-701 Korea 
\\~\\
email: taejin@kangwon.ac.kr}

\maketitle

\centerline{\bf Astract}

We study a spin dependent Tomonaga-Luttinger model in one dimension, which describes electron transport through a single barrier. Using the Fermi-Bose equivalence in one dimension, 
we map the model onto a massless Thirring model with a boundary interaction. 
A field theoretical perturbation theory for the model has been developed and the chiral symmetry is found to play an important role. The classical bulk action possesses a global $U_A(1)^4$ chiral symmetry, since the fermion fields are massless. This global chiral symmetry is broken by the boundary interaction and the bosonic degrees of freedom, corresponding to the chiral phase 
transformation, become dynamical. They acquire an additional kinetic action from the fermion path integral measure
and govern the critical behaviors of physical operators. On the critical line where the boundary interaction becomes marginal, they decouple from the fermi fields. Consequently the action reduces to 
the free field action, which contains only a fermion bilinear boundary mass term as an interaction term.  
By a renormalization group analysis, 
we obtain a new critical line, which differs from the previously known critical lines in the literature. 
The result of this work implies that the phase diagram of the one dimensional electron system may have a richer structure than previously known.  



\vskip 2cm

\section{Introduction}

The chiral symmetry \cite{Bell,Adler} has been a key ingredient to understand many important subjects in particle physics and mathematical physics such as the neutral pion decay \cite{Bell}, anomaly cancellation conditions in gauge theories \cite{Bouchiat,Nieuwenhuizen}, and instanton physics \cite{Schwarz,Atiyah}. 
We may enlarge the scope of its applications to condensed matter physics, which offers an exciting playground for both theorists and experimentalists with various different backgrounds. 
The subject, to be discussed in this paper is the $U(1)$ chiral symmetry in a 
field theoretical model, which describes electron transport through a single barrier in $1+1$ dimensions. 

The model is formulated 
initially in terms of boson fields, representing the charge and spin degrees of freedom, as a spin dependent Tomogana-Luttinger model \cite{Tomonaga,Luttinger} with a single barrier. It is this boson model, which has been studied extensively in the literature \cite{kane:1992a,kane:1992b,furusaki}. 
By virtue of Fermi-Bose equivalence in one dimension, boson fields can be mapped into fermion fields and 
vice versa. So it is possible to discuss the same model in two different frameworks. 
Although the model in both frameworks are equivalent to each other, often some features of the model are seen more transparently in one framework than the other. The critical behaviors of the Tomonaga-Luttinger model,
which we are about to discuss in this paper, may be the case. At the critical points the boundary interaction, which is the only interaction at the critical point, can be represented by a bilinear operator in terms of fermion fields. In the framework of fermion theory, it is easy to understand that the models are exactly soluble and all the radiative corrections vanish at the critical points. 
Thus, the fermion theory may offer more adequate framework to explore the critical behaviors of the one dimensional electron system.

The electron transport in one dimension through a single barrier may be described by the Luttinger liquid model with a periodic boundary potential \cite{kane:1992b}. The model is formally 
equivalent to a quantum dissipative system subject to a periodic potential of Caldeira and Leggett type \cite{caldeira83ann}, called the Schmid model \cite{schmid,guinea}. The electronic transport of the one dimensional
system has a quite interesting feature since the model exhibits phase transitions just as the Schmid model does. Introducing one more boson field corresponding to the spin density wave, one can describe the transportation of the electron with spin \cite{furusaki}.
The main purpose of this work is to develop a field theoretical perturbation theory for the model
and to explore their critical behaviors in the scheme of perturbation theory. 
To this end we choose the fermion framework.

In order to fermionize the model we may introduce two more auxiliary boson fields, just as we introduce an 
auxiliary boson field to fermionize the Schmid model. These auxiliary boson fields satisfy the Dirichlet boundary condition, so that they do not appear on the boundary action. Then by the Fermi-Bose equivalence,
the bulk action of the four boson fields is transcribed into 
the self-interacting massless Thirring action with four flavors and the boundary periodic potential into the 
boundary fermion mass term. Since the bulk action does not contains a fermion mass term, the bulk action is 
invariant under the global $U_A(1)^4$ chiral and $U(1)_V^4$ vector phase transformations. 
But $U(1)_A^4$ chiral symmetry is broken by the boundary interaction and corresponding 
boson degrees of freedom become dynamical. 
The role of the boson fields corresponding to the local $U(1)_V^4$ vector phases of the fermi fields is
insignificant. Since they are free boson fields, which do not couple to any physical operator, 
they can be safely removed from the action.
In contrast, the boson fields, which correspond to the local $U(1)_A^4$ chiral phase paly an important role. Their bulk action receives additional contribution from the path integral measure through the chiral anomaly and they couple to the fermi fields. The radiative corrections to the boundary interaction mainly arise from the interaction between these boson fields and the fermi fields of charge and spin degrees of freedom. It is these boson fields which govern the critical behaviors of physical operators. If the contributions of the chiral anomaly to their kinetic action were not taken into account correctly, the renormalization group analysis of physical operators would be erroneous. At the critical points the boson fields of the local chiral phase of the fermi fields, 
decouple from the fermi fields and the action for the fermi fields becomes a free field one with a bilinear boundary mass term only.

\section{Chiral $U_A(1)$ Symmetry:\\ Transport of Spinless Electrons}

We begin with a simpler model of the transport of spinless electrons through a single barrier first. 
The model is described by a single boson field and a periodic boundary potential
\beq \label{sine1}
S = \frac{\a}{4\pi}
\int d\t d\s\, \left(\p_\t \phi \p_\t \phi + \p_\s \phi \p_\s \phi\right)  + \frac{V_0}{2\pi}
\int d\t \left(e^{i\phi}+ e^{-i\phi} \right)\Bigl\vert_{\s =0}.
\eeq
If we integrate out the boson field $\phi$ on the bulk,
we get a non-local boundary action,
\beq \label{schmid}
S = \frac{\eta}{4\pi}\int d\tau d\tau^\prime \,
\frac{\left(\phi(\t) - \phi(\t^\prime)\right)^2}{(\t-\t^\prime)^2} 
+ \frac{V_0}{2\pi} \int d\tau \left(e^{i\phi} + e^{-i\phi} \right)\Bigl\vert_{\s =0}. 
\eeq 
where $\eta = \a/(2\pi)$. This model is known as the Schmid model, which has been studied extensively in the literature \cite{guinea,fisher;1985} since the seminal paper by Schmid \cite{schmid}. 
The model depicts a quantum dissipative 
system of Caldeira and Leggett type subject to a periodic potential and the parameter $\eta$ in Eq.(\ref{schmid})
is the friction coefficient. 
This action can be interpreted also as the open string action subject to a boundary periodic potential. 
The parameter $\a$ in Eq.(\ref{sine1}) then can be interpreted the inverse of the Regge slope $\ap$, which is related to the tension $T$ of the string as follows
\beq
T = \frac{1}{2\pi \ap} = \frac{\a}{2\pi} .
\eeq
When $\a= 1$, the system becomes critical. The action at the critical point has been studied in great 
detail in string theory because the model is proposed to depict the decay of unstable D-branes.
It is called the rolling tachyon \cite{Sen:2002nu,senreview,Lee:2005ge,Hassel}. The critical point $\a=1$ is also the self-dual point under the 
particle-kink duality \cite{schmid}. 

\subsection{Fermionization of the Model for Spinless Electrons}

It is easy to understand why the system becomes critical at the point, $\a=1$ if we
fermionize the model. It is straightforward to transcribe the bulk boson action into a bulk fermion 
action by using the well-established Fermi-Bose equivalence. But we must elaborate further in order to 
fermionize the boundary interaction term $:e^{i\phi}:$. The fermi fields
$\psi_L$ and $\psi_R$ are represented in terms of the boson fields as $:\eta_L e^{-i\sqrt{2} \phi_L}:$
and $:\eta_R e^{i\sqrt{2} \phi_R}:$ repectively where $\eta_{L/R}$ are Klein factors. At a glance we find 
that the boundary term $:e^{i\phi}:$ cannot be written as a fermion bilinear operator like
$\psi^\dag_L \psi_R$. This problem can be resolved \cite{Polchinski:1994my} 
by introducing an auxiliary boson field $\varphi$, which satisfies the Dirichlet 
condition $\varphi = 0$ at the boundary. Then the composite operator $:e^{i(\phi\pm\varphi)}:$  
has precisely the property that we want. Defining two boson fields 
$\Phi_1 = (\varphi+\phi)/\sqrt{2}, ~~\Phi_2 =(\varphi-\phi)/\sqrt{2} $, 
we may write the composite operator $:e^{i(\phi\pm\varphi)}:$ 
as $:e^{\pm i\sqrt{2} \Phi_i}:$, $i=1, 2$, which can be represented as a fermion bilinear operator 
both in the bulk and on the boundary in a consistent way. 
Because of the Dirichlet boundary condition,
$\varphi =0$ on the boundary, the composite  $:e^{i(\phi\pm\varphi)}:$ coincides
with the boundary potential, $:e^{i\phi}:$ on the boundary.

In terms of two boson fields $\Phi_i$, $i=1,2$, we may rewrite the action Eq.(\ref{sine1}) as 
\beq \label{sine2}
S =  \frac{\a}{4\pi}\int d\t d\s \sum_i^2 \p \Phi_i \p \Phi_i
+ \frac{V_0}{4\pi} \int d\s \sum_i^2 \left(e^{i\sqrt{2}\Phi_i}+ e^{-i\sqrt{2} \Phi_i}\right)\Biggl\vert_{\t=0}.
\eeq
It should be noted that in Eq.(\ref{sine2}) the two dimensional space-time coordinates $\t$ and $\s$ are interchanged, since it is more convenient to work in the closed string picture. Hereafter we will
work on the model defined in the closed string picture, where the boundary is the spatial line, $\t=0$. 
The auxiliary boson field $\varphi$ is a free boson field in the bulk and vanishes on the boundary.
It is completely decoupled from the physical degrees of freedoms and its role is to give the right conformal dimension to the boundary operator. 
The boson fields $\Phi_{i L/R}$, $i=1,2$ are mapped onto the fermion fields as 
\begin{subequations}
\beq
\psi^1_L &=& e^{-\frac{\pi}{2} i \left(p^1_L +2 p^2_L + p^1_R + 2p^2_R\right)} e^{-\sqrt{2} i \Phi^1_L},\\
\psi^2_L &=& e^{-\frac{\pi}{2} i \left(p^2_L+ p^2_R\right)} e^{-\sqrt{2} i \Phi^2_L},\\
\psi^1_R &=& e^{-\frac{\pi}{2} i \left(p^1_L +2 p^2_L + p^1_R + 2p^2_R\right)} e^{\sqrt{2} i \Phi^1_R} \\
\psi^2_R &=& e^{-\frac{\pi}{2} i \left(p^2_L+ p^2_R\right)} e^{\sqrt{2} i \Phi^2_R}.
\eeq
\end{subequations}
Here we give an explicit expression of the Klein factors, which ensure the anti-commutation relation between the fermi fields. The Klein factors are not unique. One may find some other, yet equivalent representations of Klein factors.  

At the critical point $\a=1$, the bulk action in the fermion theory is the 
free fermion action with two flavors 
and the boundary periodic interaction terms are replaced by fermion boundary mass terms, which are only quadratic in the fermion fields.
\beq \label{fermion1}
S &=& \frac{1}{2\pi}\int d\tau d\s ~
\sum_{i=1}^2 \bar\psi_i\g^\m \p_\m \psi_i + \frac{V_0}{4\pi} \int d\s \sum_{i=1}^2 \bar\psi_i \psi_i \Bigl\vert_{\t=0} ,
\eeq
where $\psi_i = (\psi_{iL}, \psi_{iR})^t$, $i=1, 2$ and 
\beq
\g^0 = \s_1 = \left(\begin{array}{cc}
  0 & 1 \\
  1 & 0 
\end{array}\right), ~~
\g^1 = \s_2= \left(\begin{array}{cc}
  0 & -i \\
  i & 0 
\end{array}\right), ~~ \g^5 = -i \g^0 \g^1= \s_3 = \begin{pmatrix} 1 & 0 \\ 0 & -1 \end{pmatrix}.
\eeq
The action of the fermion model Eq.(\ref{fermion1}) clearly shows that the model is exactly solvable and the 
boundary interaction terms do not receive any radiative correction at the critical point. 

At the off-critical point where $\a \not =1$, we may write the action as
\beq
S &=&  \frac{1}{4\pi}\int d\t d\s \sum_i^2 \p \Phi_i \p \Phi_i
+ \frac{\a-1}{4\pi}\int d\t d\s \sum_i^2 \p \Phi_i \p \Phi_i \nn\\
&& + \frac{V_0}{4\pi} \int d\t \sum_i^2 \left(e^{i\sqrt{2}\Phi_i}+ e^{-i\sqrt{2} \Phi_i}\right)\Bigl \vert_{\t=0},
\eeq
and treat the second term as a part of the interaction action. By the Fermi-Bose equivalence 
\begin{subequations}
\beq
j^0_i &=& \bar\psi_i \g^0 \psi_i = \psi^\dagger_{iL} \psi_{iL} + \psi^\dagger_{iR}\psi_{iR} 
= \sqrt{2} \p_\s \Phi_i, \\
j^1_i &=& \bar\psi_i \gamma^1 \psi_i = i
\psi^\dagger_{iL} \psi_{iL} - i \psi^\dagger_{iR} \psi_{iR} = -\sqrt{2} \p_\t \Phi_i, ~~ i= 1, 2 ,
\eeq
\end{subequations}
the second term is transcribed into the Thirring interaction \cite{Thirring} term in the fermion theory
Hence, the fermionized action is given by
\beq
S &=&  \frac{1}{2\pi} \int d\t d\s \sum_i^2
\left(\bar{\psi}_i \gamma^\m \p_\m 
\psi_i + \frac{g}{4} j^\m_i j_{i\m}\right) 
+ \frac{V_0}{4\pi} \int d\s \sum_i^2 {\bar \psi}_i \psi_i \Bigl\vert_{\t=0}
\eeq
where $g = \a -1$.     

Introducing Abelian vector fields $A_{i\m}$, $i=1,2$, we may rewrite the fermion action as follows
\beq
S &=& \frac{1}{2\pi} \int d\t d\s\sum_i^2
\left[
\bar{\psi}_i\gamma^\m\left(\p_\m + iA_{i\m}\right)\psi_i 
+ \frac{1}{g} A_{i\m} A_i{}^\m \right] 
+  \frac{V_0}{4\pi} \int d\s \sum_i^2 {\bar \psi}_i \psi_i \Bigl\vert_{\t=0}.
\eeq
Since the Abelian vector fields in $1+1$ dimensions are generally decomposed as 
\beq
A^\m_i = \e^{\m\n} \p_\n \th_i + \p^\m \chi_i, ~~~ i =1,2 ,
\eeq
we may rewrite the bulk action also in the following form
\beq \label{bulk3}
S_{\rm bulk} = \frac{1}{2\pi} \int d\t d\s\sum_i^2
\left[
\bar{\psi}_i\gamma^\m\left(\p_\m + i\e^{\m\n} \p_\n \th_i + i \p^\m \chi_i
\right)\psi_i 
+ \frac{1}{g}\left(\p\th_i \p\th_i+ \p\chi_i \p\chi_i\right)  \right].
\eeq
The interactions between the boson fields, $\th_i$, $\chi_i$ and the fermion fields in the bulk action
Eq.(\ref{bulk3})
are removed by a local $U_V(1)^2 \times U_A(1)^2$ phase transformation
\beq
\psi_i = e^{-i\g_5 \th_i -i\chi_i} \psi_{i\,0},\quad 
\bar\psi_i = \bar\psi_{i\,0} e^{-i\g_5 \th_i +i\chi_i}.
\eeq
Then the bulk action is transformed into a free field action 
\beq \label{bulk2}
S_{\rm bulk} = \frac{1}{2\pi} \int d\t d\s \sum_{i=1}^2
\left[
\bar\psi_{i\,0} \gamma^\m \p_\m \psi_{i\,0} + \frac{1}{g} \left(\p\th_i \p\th_i+ 
\p\chi_i \p\chi_i\right) \right].
\eeq
If the boson fields $\th_i$ and $\chi_i$, corresponding to the local chiral and vector phases of the fermion fields respectively, are not coupled to physical operators in the boundary action, 
we may drop them from the action. 

The local $U_A(1)^2$ chiral symmetry is broken in two ways: It is broken by 
the boundary interaction since the 
boundary mass term transforms under the chiral phase transformation as   
\beq
\sum_i \bar\psi_i\psi_i = \sum_i \bar\psi_{i0} e^{-2i \g_5 \th_i} \psi_{i0} \,.
\eeq           
And it is also broken by the chiral anomaly, which is manifested as the non-invariance of the path integral measure under the chiral transformation \cite{Fujikawa,Fujikawa2004,Naon}
\beq
D[\psi]D[\bar\psi] = D[\psi_0] D[\bar\psi_0] \exp
\left[-\frac{1}{2\pi} \int d\t d\s \sum_i (\p \th_i)^2 \right].
\eeq
It yields an additional kinetic action for $\th_i$ in the phase transformed action Eq.(\ref{bulk2}) so that 
the action $S$ is written as 
\beq
S &=& \frac{1}{2\pi} \int d\t d\s \sum_{i=1}^2
\left[
\bar\psi_{i} \gamma^\m \p_\m \psi_{i} + \left(\frac{1+g}{g}\right) \p\th_i \p\th_i+ 
\p\chi_i \p\chi_i \right] \nn\\
&& + \frac{V_0}{4\pi} \int d\s \sum_i^2 \bar\psi_{i} e^{-2i \g_5 \th_i} \psi_{i}\Bigl\vert_{\t=0}
\eeq
where we omit the subscript $0$ for the fermi fields 
for the sake of convenience. While the $U(1)$ chiral symmetry is broken, the $U(1)$ vector symmetry
$U_V(1)^2$ is kept unbroken: The measure and the boundary interaction are invariant under the vector phase transformation. Since the vector phase transformation is realized for the boson fields $\chi_i$ as a translation and $\chi_i$ are not coupled to the physical fermion fields, we may drop them from the 
action. 

The final step is to scale the boson fields $\th_i$ as 
\beq
\th_i \rightarrow \kappa \th_i, ~~~
\kappa = \sqrt{\frac{g}{2(1+g)}} = \sqrt{\frac{\a-1}{2\a}},\, ~~ i = 1, 2.
\eeq
It brings us to the following action, which contains a boundary interaction only
\beq \label{action3}
S = \frac{1}{2\pi} \int d\t d\s \sum_{i=1}^2
\left\{
\bar\psi_{i} \gamma^\m \p_\m \psi_{i} + \half\left(\p\th_i\right)^2 \right\} + \frac{V_0}{4\pi} \int d\s \sum_{i=1}^2 \bar\psi_{i} e^{-2i\g_5 \kappa\th_i} \psi_{i}\Bigl\vert_{\t=0} \,.
\eeq
We note that the obtained action has a considerable advantage, being compared to other forms of the action.
The bulk action contains only the free field action for fermion fields and boson fields 
and all the non-trivial interactions are encoded in the boundary action. Therefore, we only need to 
deal with the correlation functions on the one dimensional boundary. The only role of the bulk action
is to define the free fermion and boson Green's functions on the boundary. 
From Eq.(\ref{action3}) it is clear that in the limit of the critical point, $\kappa \rightarrow 0$ ($\a \rightarrow 0$), the coupling between the boson fields $\th_i$ and the fermion fields $\psi_i$ vanishes. 
Thus, $\th_i$, becoming free fields, can be dropped from the action and the action reduces to the 
free fermion action  with a boundary mass. 
In the next section we shall calculate the radiative corrections to the boundary interaction term, using the 
obtained action Eq.(\ref{action3}). 

\subsection{Radiative Corrections to the Boundary Periodic Potential}

Our discussion on the radiative corrections begin with an expansion of the boundary action in ${\kappa}$, 
which is a small parameter near the critical point
\beq
S_{\rm boundary} = \frac{V_0}{4\pi} \int d\s \sum_{i} \left[\bar\psi_{i}\psi_{i}
-2 i \bar\psi_{i} \g_5 \kappa\th_{i} \psi_{i} 
-2 \bar\psi_{ia} \left(\g_5 \kappa\th_{i}\right)^2 \psi_{i} + \cdots
\right] .  
\eeq      
In order to calculate the radiate corrections to the boundary periodic potential we consider 
the fully interacting two-point function (Green's function) on the boundary \cite{brown}. The fully interacting Green's function in the bulk (on the cylinder) is defined as 
\beq
\bolG_{(i|\a\b)}(\t_1-\t_2;\s_1-\s_2) &=&
\langle 0 \vert T \psi_{i\a}(\t_1,\s_1) \barpsi_{i\b}(\t_2,\s_2) \vert 0\rangle \nn\\
&=& \int D[\barpsi, \psi] 
\psi_{i\a}(\t_1,\s_1) \barpsi_{i\b}(\t_2,\s_2) e^{-S_0 - S_{\rm boundary} }, 
\eeq
where 
\begin{subequations}
\beq
S_0 &=& \frac{1}{2\pi} \int d\t d\s \sum_{i=1}^2\left[
\bar\psi_{i} \gamma^\m \p_\m \psi_{i} + \half\left(\p\th_i\right)^2 \right], \\
S_{\rm boundary} &=& \frac{V_0}{4\pi} \int d\s \sum_{i=1}^2 \bar\psi_{i} e^{-2i\g_5 \kappa\th_i} \psi_{i}\Bigl\vert_{\t=0} .
\eeq
\end{subequations}
And the fully interacting Green's function on the boundary $\bolG(\s_1-\s_2)$ is defined as 
the bulk Green's function evaluated on the boundary
\beq \label{limit}
\bolG_{\a\b}(\s_1-\s_2) &=& \lim_{\t_1, \t_2 \rightarrow 0} 
\langle 0 \vert T \psi_{\a}(\t_1,\s_1) \barpsi_{\b}(\t_2,\s_2) \vert 0\rangle .
\eeq
Since the action is diagonal in the flavor index $i$, the correlation functions, to be calculated, do not
depend on the flavor index $i$. Hereafter the flavor index $i$ shall be omitted for notational convenience. We also leave out the coordinate $\t$, since only operators on the boundary concern us.  

For the Fermi fields on a cylinder, two conditions are available. We may choose the anti-periodic condition
\beq
\psi (2\pi) = -\psi (0), ~~ \barpsi (2\pi) = -\barpsi (0),~~
\xi (2\pi) = -\xi (0), ~~ \barxi (2\pi) = -\barxi (0),
\eeq
or the periodic condition
\beq
\psi (2\pi) = \psi (0), ~~ \barpsi (2\pi) = \barpsi (0),~~
\xi (2\pi) = \xi (0), ~~ \barxi (2\pi) = \barxi (0).
\eeq
The former is called the Neveu-Schwarz (NS) sector where the normal modes are labeled by half-odd-integers,
$n \in \bbZ + 1/2$
and the latter is called the Ramond (R) sector where the normal modes are labeled by integers, 
$n \in \bbZ$. Here we choose the NS sector only, because the physical vacuum belongs to the NS sector, of 
which vacuum does not carry non-vanishing momentum. The vacuum in the R sector carries a
non-vanishing momentum. 

The Green's function on the boundary at the lowest order is defined as the free Green's function evaluated on the boundary, $G(0;\s_1-\s_2)$. 
The fermion Green's function on the boundary has two different Fourier decompositions as follows, depending 
the direction, along which the limit Eq.(\ref{limit}) is taken
\begin{subequations}
\beq
G(0+;\s_1-\s_2) &=& \sum_{n \in \bbZ+1/2} \frac{1}{2\pi} \g^0  \begin{pmatrix}
\th(n) & 0 \\ 0 & \th(-n) \end{pmatrix} e^{in(\s_1-\s_2)} , \\
G(0-;\s_1-\s_2) &=& -\sum_{n \in \bbZ+1/2} \frac{1}{2\pi} \g^0  \begin{pmatrix}
\th(-n) & 0 \\ 0 & \th(n) \end{pmatrix} e^{in(\s_1-\s_2)} , 
\eeq
\end{subequations}
where $\th(n)$ is the unit step function (or Heaviside step function),
\beq
\th(n) = \begin{cases}  1 & \quad \text{for}~~ n \ge 0 \\
0 & \quad \text{for}~~ n < 0 ~~.
\end{cases} 
\eeq
Although they yield the same Green's function $G(\s)$ of closed form given as 
\beq
G(\s) = \g^0 \frac{i}{4\pi \sin(\s/2)} \begin{pmatrix} 1& 0 \\ 0 & -1 \end{pmatrix},
\eeq
we must be careful when choosing the Fourier decomposition in the calculations of correlation
functions.  

At tree level the Green's function on the boundary receives a correction 
from the boundary interaction
\beq
\bolG^{(0)}_{\a\b} (\s_1-\s_2) &=& \frac{1}{2!} \left(\frac{V_0}{2}\right)^2 
\Biggl\langle\psi_{\a}(\s_1) \barpsi_{\b}(\s_2) 
\left(\int\frac{d\s}{2\pi}\bar\psi \psi \right)^2
\Biggr\rangle^{(0)} \nn\\
&=& \frac{V_0^2}{4}\int\frac{d\sp}{2\pi}\int\frac{d\spp}{2\pi} \Bigl(G(\s_1-\sp) G(\sp-\spp) G(\spp-\s_2)\Bigr)_{\a\b}   .
\eeq
Recall that we treat the boundary mass term as a part of interaction 
even though the boundary mass term is a bilinear operator in terms of the fermi fields.
Defining the Fourier transformation of the Green's functions as 
\beq \label{green1}
\bolG_{\a\b}^{(0)}(\s) = \sum_{n\in \bbZ+1/2} \bolG_{\a\b}^{(0)}[n] e^{in\s},~~
G(\s) = \sum_{n\in \bbZ+1/2} G[n] e^{in\s}, 
\eeq
we find 
\beq \label{green2}
\bolG^{(0)}[n] = \frac{V_0^2}{4} \frac{1}{(2\pi)^2}  G[n] .
\eeq

At the first order the one-loop correction to the Green's function is given as 
\beq
\bolG^{(1)}_{\a\b} (\s_1-\s_2) &=& \frac{V_0^2}{2!} \Biggl\langle\psi_{\a}(\s_1) \barpsi_{\b}(\s_2) 
\left(\int\frac{d\s}{2\pi}\bar\psi \g_5 \kappa\th \psi \right)^2
\Biggr\rangle^{(0)} \nn\\
&=& V_0^2 \k^2 \int\frac{d\sp}{2\pi}\int\frac{d\spp}{2\pi} \nn\\
&& \Bigl(G(\s_1-\sp)\g^5 G(\sp-\spp) \g^5 G(\spp-\s_2)\Bigr)_{\a\b} G_\th(\sp-\spp)  .
\eeq
Here $G_\th(\sp-\spp)$ is the Green's function of the boson field $\th$ on the boundary, which defined as
\beq
G_\th(\sp-\spp) = \langle \th(0,\sp)\th(0,\spp)\rangle^{(0)} = \sum_{n\in \bbZ} \frac{1}{2|n|} e^{in(\sp-\spp)} .
\eeq

\begin{figure}[htbp]
   \begin {center}
    \epsfxsize=0.7\hsize
%
	\epsfbox{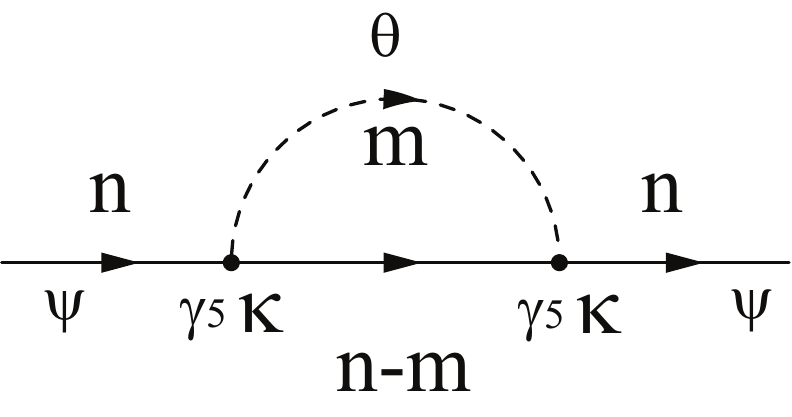}
   \end {center}
   \caption {\label{sunset1} The first order correction to the Green's function of spinless Electron}
\end{figure}

The Feynman diagram Fig.1. depicts the evaluation of the first order correction to the Green's function of 
spinless electron. 
Since the Fourier transformation of a convolution is the simple product of the Fourier transformed terms,
the Fourier transformation of the first-order Green's function may be written as 
\beq
\bolG^{(1)}[n] = V_0^2 \kappa^2 G[n] \left\{
\sum_{m \in \bbZ+1/2} \g^5 G[m] \g^5 G_\th[n-m] \right\} G[n] ,
\eeq
where $\bolG_{\a\b}^{(1)}[n]$ and $ G_\th[n]$ are the Fourier transformation of the Green's functions 
defined as 
\beq
\bolG_{\a\b}^{(1)}(\s) = \sum_{n\in \bbZ+1/2} \bolG_{\a\b}^{(1)}[n] e^{in\s},~~
G_\th(\s) = \sum_{n\in \bbZ} G_\th[n] e^{in\s}  . 
\eeq

By some algebra, we find 
\beq
\bolG^{(1)}[n] &=&  \frac{V_0^2 \k^2}{4\pi} \sum_m \frac{\g^0}{2\pi} \begin{pmatrix} \th(n)\th(-m) & 0 \\ 0 & \th(-n)\th(m)
\end{pmatrix} \frac{1}{|n-m|} \nn
\eeq
This one-loop correction is divergent as expected. We must regularize it. Note that the two diagonal components of $\g^0 \bolG^{(1)}[n]$ are rewritten as 
\begin{subequations}
\beq
\sum_{m\in \bbZ+1/2} \th(n)\th(-m) \frac{1}{|n-m|} &=& \th(n) \sum_{m=1/2}^\infty  \frac{1}{|n+m|}\nn \\
&=& \th(n) \zeta(1) + \text{finite terms}, \\
\sum_{m\in \bbZ+1/2} \th(-n)\th(m) \frac{1}{|n-m|} &=& \th(-n) \sum_{m=1/2}^\infty \frac{1}{||n|+m|}\nn\\
&=& \th(-n) \zeta(1) + \text{finite terms} 
\eeq
\end{subequations}
and they can be regularized by the Riemann zeta function $\zeta(s) = \sum_{n=1}^\infty \frac{1}{n^s}$, 
\beq
\zeta (1) = 1 + \frac{1}{2}+ \frac{1}{3} + \frac{1}{4} + \cdots = \lim_{\Lambda \rightarrow \infty}
\left( \gamma + 2 \ln \Lambda^2 \right),
\eeq
where $\g$ is the Euler–Mascheroni constant, $\g = 0.57721566 \dots$.
In passing we also note that this divergence corresponds to the ultraviolet divergence and the one-loop correction to the Green's function is free of infrared divergence.

Up to the one-loop order the radiative corrections to the Green's function are evaluated in the momentum number space as 
\beq
\bolG^{(0)}[n] + \bolG^{(1)}[n] = \frac{V^2_0}{16\pi^2} 
\left(1 + \frac{\k^2}{2} \zeta(1) + \cdots \right) G[n].
\eeq
This divergence can be removed by renormalizing the coupling constant $V$ as follows
\beq
V^2 = {V^2_0} \left( 1 + \k^2 \ln \frac{\Lambda^2}{\mu^2}\right).
\eeq 
This renormalization group flow defines the phase structure of the model,
\beq
V^2_0 \left( 1 + \frac{\a-1}{2\a} \ln \frac{\Lambda^2}{\mu^2}\right)= 
V_0^2 \left(\frac{\Lambda^2}{\mu^2}\right)^{\frac{\a-1}{2\a}} .
\eeq
When $\a>1$ the periodic potential becomes a relevant operator and when $\a <1$ 
the potential becomes an irrelevant operator. Thus, in the region where $\a>1$, the 
periodic potential is strong and the particles are mostly localized in the 
minima of the potential. In the other region where $\a<1$, the potential is weak and 
particles are delocalized. The point $\a=1$ sets the phase boundary. 
The higher order corrections may generate descendent perturbations of type $e^{in\phi}$, $n=2 ,3 , \dots$,
but they do not change the phase structure of the model. The perturbation analysis shows that 
the operator $e^{in\phi}$ becomes a relevant operator where $\a = n^2 >1$. In this region
the particles are already localized by the periodic potential due to the primary operator $e^{i\phi}$.

\section{Chiral $U_A(1) $ Symmetry in \\
Transport of Electrons with Spin} \label{sectionspin}

Being equipped with the field theoretical perturbation tool, developed in the last section,
we will discuss the more complex system, namely the 
transport of one-dimensional electrons with spin. The one-dimensional single-channel interacting electron with spin system is described by the following action \cite{furusaki}
\beq \label{spin}
S &=& \frac{1}{4\pi} \int^{\beta_T}_0 dt \int dx\, \left[\frac{1}{v_\rho \eta_\rho}\left(\p_t \th\right)^2 + \frac{v_\rho}{\eta_\rho}\left(\p_x \th\right)^2 + \frac{1}{v_\s \eta_\s}\left(\p_t \phi\right)^2 + \frac{v_\s}{\eta_\s}\left(\p_x \phi\right)^2 \right]\\
&& ~ +V_0^\prime \int^{\beta_T}_0 dt \cos \th \cos \phi \Bigl\vert_{x=0}\nn
\eeq
where the phase fields $\th$ and $\phi$ represent charge and spin density fluctuations respectively and the boundary term represents the impurity (barrier) potential. We refer the reader to ref.\cite{furusaki} for parameters
appearing in Eq.(\ref{spin}).  
We may map the model into a string theory action with a periodic tachyon potential on a disk
\beq
S &=& \frac{\a_\rho}{4\pi} \int d\t d\s \left[(\p_\t \th)^2 + (\p_\s \th)^2 \right]
+ \frac{\a_\s}{4\pi} \int d\t d\s \left[(\p_\t \phi)^2 + (\p_\s \phi)^2 \right] \nn\\
&& + 2V_0 \int \frac{d\s}{2\pi} \cos\th \cos\phi \Bigl\vert_{\t=0}
\eeq
where 
\beq
\s = \frac{2\pi}{\b_T}\, t, ~~~ \a_\rho = \frac{1}{\eta_\rho}, ~~~
\a_\s = \frac{1}{\eta_\s},~~~
V_0 = \frac{\b_T}{2} V_0^\prime
\eeq
and 
\beq
\th(\t,\s) = \th(\frac{2\pi}{\b_T} \frac{1}{v_\rho}\, x, \frac{2\pi}{\b_T}\, t),
~~~
\phi(\t,\s) = \phi(\frac{2\pi}{\b_T} \frac{1}{v_\s}\, x, \frac{2\pi}{\b_T}\, t)
\eeq
Since the action is given as a direct sum of the actions for $\th$ and $\phi$, we scale $x$ 
differently in their actions. (Note that the boundary interaction term does not depend on $\tau$.)

It is convenient to rewrite the action in terms of scalar fields $\phi_a$, $a =1,2$ defined as 
\beq
\phi_1 = \th+ \phi, ~~~ \phi_2 = \th- \phi.
\eeq
Then the action is written as 
\beq
S_\phi = \frac{1}{4\pi} \int d\t d\s \left[\p\phi_a \p \phi_a + g^{ab} \p \phi_a \p \phi_b \right] +
\frac{V_0}{2} \int \frac{d\s}{2\pi} \sum_{a=1}^2 \left(e^{i\phi_a} + e^{-i\phi_a}\right)\Bigl\vert_{\t=0}
\eeq
where $g^{11} = g^{22} = \frac{\a_\rho+\a_\s}{4} -1,~ g^{12} = g^{21} = \frac{\a_\rho-\a_s}{4}$. If $g^{ab}=0$, or equivalently, $\a_\rho= \a_\s = 2$ ($\eta_\rho = \eta_\s = 1/2$), we see that the action reduces to a critical theory with a sum of two commuting marginal perturbations
\beq
S_\phi = \frac{1}{4\pi} \int d\t d\s\,\p \phi_a \p \phi_a + \frac{V_0}{2}\int \frac{d\s}{2\pi} \sum_a \left(
e^{i\phi_a} + e^{-i\phi_a} \right)\Bigl\vert_{\t=0}.
\eeq
So it is clear that the point $\a_\rho= \a_\s = 2$ ($\eta_\rho = \eta_\s = 1/2$) must be the critical point,
where the model can be described by four Dirac fermion fields with a bilinear boundary mass.

In order to apply the field theoretical perturbation analysis to the model, developed in the previous section, 
we should fermionize the model first. 
The fermionization begins with introducing auxiliary free boson fields $\varphi^a$ of which action has the same form as that for $\phi^a$, $a=1, 2$, 
except for the periodic boundary potential terms
\beq
S_\varphi = \frac{1}{4\pi} \int d\t d\s \left[\p \varphi_a \p \varphi_a + g^{ab} \p \varphi_a \p \varphi_b \right] .
\eeq
Here the Dirichlet boundary condition is chosen for $\varphi_a$.
Since no boundary terms for $\varphi_a$ are introduced and the bulk action for $\varphi_a$
is the free field action, $\varphi_a$ is decoupled from the physical fields.
Their role is to give the boundary operators the right dimensions in the bulk and on the boundary, 
so that the boundary operators can be transcribed into bilinear fermion operators.
 
Using the Fermi-Bose equivalence, we may write the fermion fields in terms of 
the boson fields as 
\beq
\psi_{a1L} &=& \zeta_{a1L} :e^{-i\sqrt{2} \Phi_{a1L}}:, ~~~
\psi_{a2L} = \zeta_{a2L} :e^{-i\sqrt{2} \Phi_{a2L}}: \\
\psi_{a1R} &=& \zeta_{a1R} :e^{i\sqrt{2} \Phi_{a1R}}:, ~~~
\psi_{a2R} = \zeta_{a2R} :e^{i\sqrt{2} \Phi_{a2R}}:, \nn
\eeq
where 
\beq
\Phi_{a1} = \frac{1}{\sqrt{2}}\left(\varphi_a+ \phi_a\right),~~~
\Phi_{a2} = \frac{1}{\sqrt{2}}\left(\varphi_a -\phi_a\right), ~~~ a =1, 2,
\eeq
and $\zeta_{aiL}$, $\zeta_{aiR}$ are cocycles which ensure the anti-commutation 
relationship between the fermion fields. 
Accordingly, the action can be written in terms of the fermion fields as 
\beq\label{fermionaction}
S &=& S_\phi + S_\varphi \nn\\
&=& \frac{1}{2\pi} \int d\t d\s \left(\sum_{a,i} \bar\psi_{ai} \g^\m \p_\m \psi_{ai} + \sum_{a,b,i}\frac{g^{ab}}{4} j^\m{}_{ai} j_\m{}_{bi} \right) + \frac{V_0}{4\pi} \int d\s \sum_{a,i} \bar\psi_{ai} \psi_{ai} \Bigl\vert_{\t=0} ~ 
\eeq
where $j_\m{}_{ai} = \bar\psi_{ai} \gamma^\m \psi_{ai}$. 
This action can be
understood as a four flavor Thirring model with boundary masses. 

We may rewrite the Thirring action as follows, introducing four Abelian $U(1)$ vector fields 
$A_{i\m a}$, $i, a = 1, 2$
\beq
S &=& \frac{1}{2\pi} \int d\t d\s \sum_{a,i} \left[\bar\psi_{ai} \g^\m \left(\p_\m +iA_{i\mu a}\right)\psi_{ai}+ A_{i\m a} (g^{-1})_{ab} A^\m_{i b}\right]  \nn\\
&&+ \frac{V_0}{4\pi} \int d\s \sum_{a,i} \bar\psi_{ai} \psi_{ai}\Bigl\vert_{\t=0}. 
\eeq
Decomposing the Abelian vector fields 
\beq
A^\m{}_{ia} = \e^{\mu\n}\p_\nu \th_{ia}+ \p^\m \chi_{ia}, \quad i=1, 2 \quad a=1, 2.
\eeq
we can remove the interaction between the vector fields and the fermion fields by a local phase transformation,
\beq
\psi_{ia} = e^{-i\g_5 \th_{ia} -i\chi_{ia}} \psi_{ia\,0},\quad 
\bar\psi_{ia} = \bar\psi_{ia\,0} e^{-i\g_5 \th_{ia} +i\chi_{ia}}.
\eeq
Under the local phase transformation the path integral measure for the fermi fields transforms as 
\beq
D[\psi]D[\bar\psi] = D[\psi_0] D[\bar\psi_0] \exp
\left[-\frac{1}{2\pi} \int d\t d\s \sum_{i,a} (\p \th_{ia})^2 \right].
\eeq
After the local phase transformation, the bulk action contains only free field actions for four fermion fields and eight boson fields 
\beq \label{bulkspin}
S_{\rm bulk} = \frac{1}{2\pi} \int d\t d\s \sum_{i=1}^2
\left[
\bar\psi_{ia\,0} \gamma^\m \p_\m \psi_{ia\,0} + \p\th_{ia}\left(g^{-1}
+I\right)_{ab} \p\th_{ib}+ \p\chi_{ia}(g^{-1})_{ab}\p\chi_{ib}
\right].
\eeq
The additional kinetic term for the boson fields $\th_{ia}$ comes from the path integral measure for the 
fermi fields. This is the manifestation of the chiral $U_A(1)^4$ anomaly. 
The action is diagonal in the flavor index $i$. If the massless boson fields $\th_{ia}$ and 
$\chi_{ia}$ do not couple to the fermion fields $\psi_{ia}$, we may drop the boson fields 
from the action, leaving the massless fermion fields only in the action. 
The global $U_A(1)^4$ symmetry is broken by the boundary interaction term
\beq
\sum_{i,\,a} \bar\psi_{ia}\psi_{ia}= \sum_{i,\,a} \bar\psi_{ia0} e^{-2i \g_5 \th_{ia}} \psi_{ia0},
\eeq
and the corresponding degrees of freedom, the boson fields $\th_{ia}$ couple to the fermi fields.
On the other hand, the boundary action is invariant under the $U_V(1)^4$ phase transformation, 
and the corresponding boson fields $\chi_{ia}$ do not 
couple to the fermi fields. Thus, we may drop these free boson fields from the action. 

We can diagonalize the matrix $g^{-1}_{ab}$ by a similarity transformation
\beq
M^{t} g^{-1} M =  \left(\begin{array}{rr}
\frac{2}{\a_\rho-2} & 0\\
0 & \frac{2}{\a_\s-2} \end{array} \right) ,
\eeq
where
\beq
M = M^{-1} = M^t = \frac{1}{\sqrt{2}} 
\left(\begin{array}{rr}
1 & 1 \\
1 & -1 \end{array} \right),\qquad  M^t M = I.
\eeq
Under the similarity transformation for $\th_{ia}=M_{ab}\th_{ia}^\prime$,
\beq \label{similarity}
g^{-1} + I \rightarrow M^{-1} g^{-1} M + I =  \left(\begin{array}{rr}
\frac{\a_\rho}{\a_\rho-2} & 0\\
0 & \frac{\a_\s}{\a_\s-2} \end{array} \right) = 
\half\left(\begin{array}{rr}
1/\kappa^2_1 & 0\\
0 & 1/\kappa^2_2 \end{array} \right) .
\eeq
The bulk action for $\th_{ia}= M_{ab} \th_{ib}^\prime$ becomes
\beq
S_{\th} = \frac{1}{4\pi}\int d\t d\s \sum_i\left[\kappa^{-2}_1 \left(\p \th^\prime_{i1}\right)^2+
\kappa^{-2}_2 \left(\p \th^\prime_{i2}\right)^2\right].
\eeq
Then we scale $\th^\prime$ to have a free field action for $\th$ in the bulk 
\beq
\th^\prime_{i1} \rightarrow \kappa_1 \th^\prime_{i1}, \quad
\th^\prime_{i2} \rightarrow \kappa_2 \th^\prime_{i2}.
\eeq
where
\beq
\kappa_1 = \sqrt{\frac{\a_\rho-2}{2\a_\rho}}, \quad
\kappa_2 = \sqrt{\frac{\a_\s-2}{2\a_\s}}.
\eeq
The scaling is equivalent to the following transformation
\beq
\th^\prime_a = {\bf K}_{ab} \th^{\prime\prime}_b,\quad {\bf K} = \left(\begin{array}{rr}
\kappa_1 & 0\\
0 & \kappa_2 \end{array} \right)
\eeq
The similarity transformation and the scaling bring us to the 
the boundary mass term 
\beq
\sum_{i,\,a} \bar\psi_{ia}\psi_{ia}= \sum_{i,\,a} \bar\psi_{ia0} e^{-2i\g_5 (M{\bf K})_{ab}\th^{\prime\prime}_{ib}} \psi_{ia0}
\eeq
and the action in the desired form
\beq
S = \frac{1}{2\pi} \int_M d\t d\s \sum_{i=1}^2\left[
\bar\psi_{ia} \gamma^\m \p_\m \psi_{ia} + \half\p\th_{ia}\p\th_{ia}\right]
+\frac{V_0}{4\pi}\int d\s \sum_{i,\,a} \bar\psi_{ia} e^{-2 i\g_5 (M{\bf K})_{ab}\th_{ib}} \psi_{ia} .
\eeq
Here we omit the subscript $0$ for the fermi fields and the superscript $\prime\prime$ for $\th$ fields
for the sake of convenience.
Near the critical line, ${\bf K}$ has small eigenvalues and the boundary action may be expanded in ${\bf K}$
\beq \label{expansion}
S_{\rm boundary} &=& \frac{V_0}{4\pi}\int d\s \sum_{i,\,a} \Bigl[\bar\psi_{ia}\psi_{ia}
-2 i \bar\psi_{ia} \g_5 (M{\bf K})_{ab}\th_{ib} \psi_{ia} \nn\\
&&~~~~~~~~~~~~~~~~~~~~~~~~~~ -2 \bar\psi_{ia} \left(\g_5 (M{\bf K})_{ab}\th_{ib}\right)^2 \psi_{ia} + \cdots
\Bigr] \Bigl\vert_{\t=0} .
\eeq
Now we are ready to calculate the radiative corrections to the boundary periodic potential in the scheme of perturbation theory of a renormalizable field theory.

\section{Radiative Corrections to the Periodic Potential}

Our discussion on the radiative corrections to the boundary periodic potential of the model of electrons with 
spin is parallel to the previous discussion on the radiative corrections in the model of spinless electrons.
Corrections to the boundary mass follow from examining the fully interacting propagator
\beq
\bolG_{(ia|\a\b)}(\s_1-\s_2) &=& 
\langle 0 \vert T \psi_{ia\a}(0,\s_1) \barpsi_{ia\b}(0,\s_2) \vert 0\rangle \nn\\
&=& \int D[\barpsi, \psi] 
\psi_{ia\a}(0,\s_1) \barpsi_{ia\b}(0,\s_2) \exp \left(-S_0- S_{\rm boundary}\right) .
\eeq
We treat the boundary mass as a part of interaction as before. The first term in the expansion of the boundary action $S_{\rm boundary}$, Eq.(\ref{expansion}) contributes to the corrections of the 
Green's function at tree level $\bolG^{(0)}_{(ia|\a\b)}$
\beq
\bolG^{(0)}_{(ia|\a\b)} (\s_1-\s_2)
&=& \frac{1}{2!}\frac{V_0^2}{4} \Biggl\langle\psi_{ia\a}(0,\s_1) \barpsi_{ia\b}(0,\s_2) 
\left(\sum_{b}\int\frac{d\s}{2\pi}\bar\psi_{ib} \psi_{ib} \right)^2\Biggl\vert_{\t=0}
\Biggr\rangle^{(0)} \nn\\
&=&  \frac{V_0^2}{4} \int\frac{d\sp}{2\pi}\int\frac{d\spp}{2\pi} \left(G(\s_1-\sp) G(\sp-\spp) G(\spp-\s_2)\right)_{\a\b}.
\eeq
Since $\bolG^{(0)}_{(ia|\a\b)} (\s)$ does not depend on the indices $i$ and $a$, we omit these indices in the following calculations.
The calculation of this tree level correction is just same as that for the model of spinless electron system 
Eqs.(\ref{green1},\ref{green2}). The 
Fourier transformation of $\bolG^{(0)}_{\a\b} (\s)$ is given by 
\beq
\bolG^{(0)}_{\a\b} [n] &=&  \int\frac{d\s}{2\pi} \bolG^{(0)}_{\a\b} (\s) e^{-in\s}=  \frac{V_0^2}{16\pi^2} G_{\a\b}[n], ~~ ~~~~~n \in \bbZ+1/2.
\eeq

The first-order correction to the  Green's function $\bolG^{(1)}_{(ia|\a\b)}$ is obtained also in a similar way.
The iteration of the second term in the expansion of the boundary action $S_{\rm boundary}$, 
Eq.(\ref{expansion}) leads to the first-order one-loop correction to the Green's function
\beq
\bolG^{(1)}_{(ia|\a\b)} (\s_1-\s_2)
&=& \frac{V_0^2}{2!} \Biggl\langle\psi_{a\a}(0,\s_1) \barpsi_{a\b}(0,\s_2) \nn\\
&&~~~
\left(\sum_{b,c}\int\frac{d\s}{2\pi}\bar\psi_{b} \g_5 (M{\bf K})_{bc}\th_{c} \psi_{b} \right)^2\Biggl\vert_{\t=0}
\Biggr\rangle^{(0)} \nn\\
&=& \frac{V_0^2}{2} \, Tr \left((M{\bf K})(M{\bf K})^t\right)\int\frac{d\sp}{2\pi}\int\frac{d\spp}{2\pi} \nn\\
&& \Bigl(G(\s_1-\sp)\g^5 G(\sp-\spp) \g^5 G(\spp-\s_2)\Bigr)_{\a\b} G_\th(\sp-\spp) .
\eeq

\begin{figure}[htbp]
   \begin {center}
    \epsfxsize=0.7\hsize
%
	\epsfbox{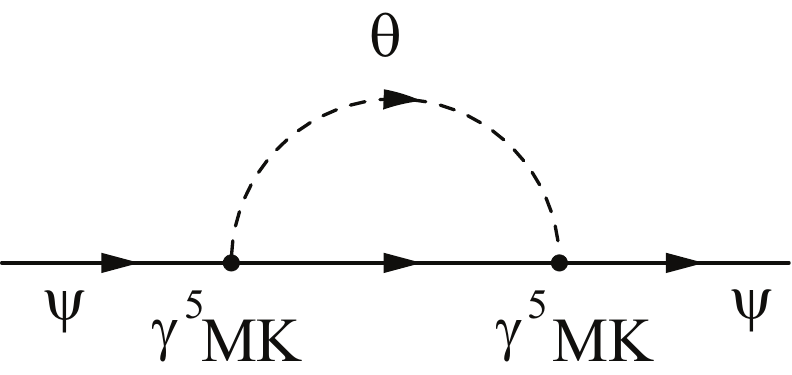}
   \end {center}
   \caption {\label{sunset2} The first order correction to the Green's function of electron with spin.}
\end{figure}

Fig 2. depicts the Feynman diagram corresponding to the first order correction $\bolG^{(1)}_{(ia|\a\b)}$.
The first order correction $\bolG^{(1)}_{(ia|\a\b)}$ is proportional to
\beq
\left({\bf M K} ({\bf MK})^t\right)_{11} &=& \left({\bf M K} ({\bf MK})^t\right)_{22} \nn\\
&=& \half( \kappa^1_1 +\kappa^2_2) \nn\\
&=& \half\left(1- \frac{1}{\a_\rho} - \frac{1}{\a_\s}\right) \nn\\
&=& \half\left(1-\eta_\rho - \eta_\s \right).
\eeq
The rest part of the calculation of $\bolG^{(1)}_{(ia|\a\b)}$ is just same as that of the first order correction to the Green's function of spinless electrons, calculated in the previous section
\beq
\bolG^{(1)}_{\a\b}[n] = \frac{V_0^2}{16\pi^2} \half \left(1- \frac{1}{\a_\rho} - \frac{1}{\a_\s}\right)\left(\zeta(1) + \text{finite terms} \right) G_{\a\b}[n],
\eeq
where $\bolG^{(1)}_{\a\b}[n]$ is the Fourier transformation of $\bolG^{(1)}_{(ia|\a\b)}$
\beq
\bolG^{(1)}_{\a\b} [n] &=&  \int\frac{d\s}{2\pi} \bolG^{(1)}_{\a\b} (\s) e^{-in\s},
~~ ~~~~~n \in \bbZ+1/2 .
\eeq

Up to the first order the radiative corrections to the Green's function is obtained 
in the momentum number space as 
\beq
\bolG^{(0)}_{\a\b} [n]+ \bolG^{(1)}_{\a\b} [n] = \frac{V^2_0}{16\pi^2}  \left(1 + \half
\left(1- \frac{1}{\a_\rho} - \frac{1}{\a_\s}\right)
\zeta(1)+\cdots \right) G_{\a\b}[n]
\eeq
As in the case of the spinless electron, we can remove this ultraviolet divergence by renormalizing the coupling constant $V$ 
\beq
V^2 &=& V^2_0 \left(1+ \left(1- \frac{1}{\a_\rho} - \frac{1}{\a_\s}\right) \ln \frac{\Lambda^2}{\m^2} \right) \nn\\
&=& V^2_0 \left(\frac{\Lambda^2}{\m^2} \right)^{1- \frac{1}{\a_\rho} - \frac{1}{\a_\s}}.
\eeq
The structure of the phase diagram is fixed by the renormalization group flow. Where 
$1- \frac{1}{\a_\rho} - \frac{1}{\a_\s} >0$ (equivalently $1-\eta_\rho-\eta_\s >0$, the region I in the figure 3) the periodic potential becomes a relevant operator and the potential tends to be strong in the zero temperature limit. In the region II
of figure 3 where $1- \frac{1}{\a_\rho} - \frac{1}{\a_\s} <0$ (equivalently $1-\eta_\rho-\eta_\s <0$), the 
periodic potential becomes an irrelevant operator and the potential is weak in the zero temperature limit.

\begin{figure}[htbp]
   \begin {center}
    \epsfxsize=0.7\hsize
%
	\epsfbox{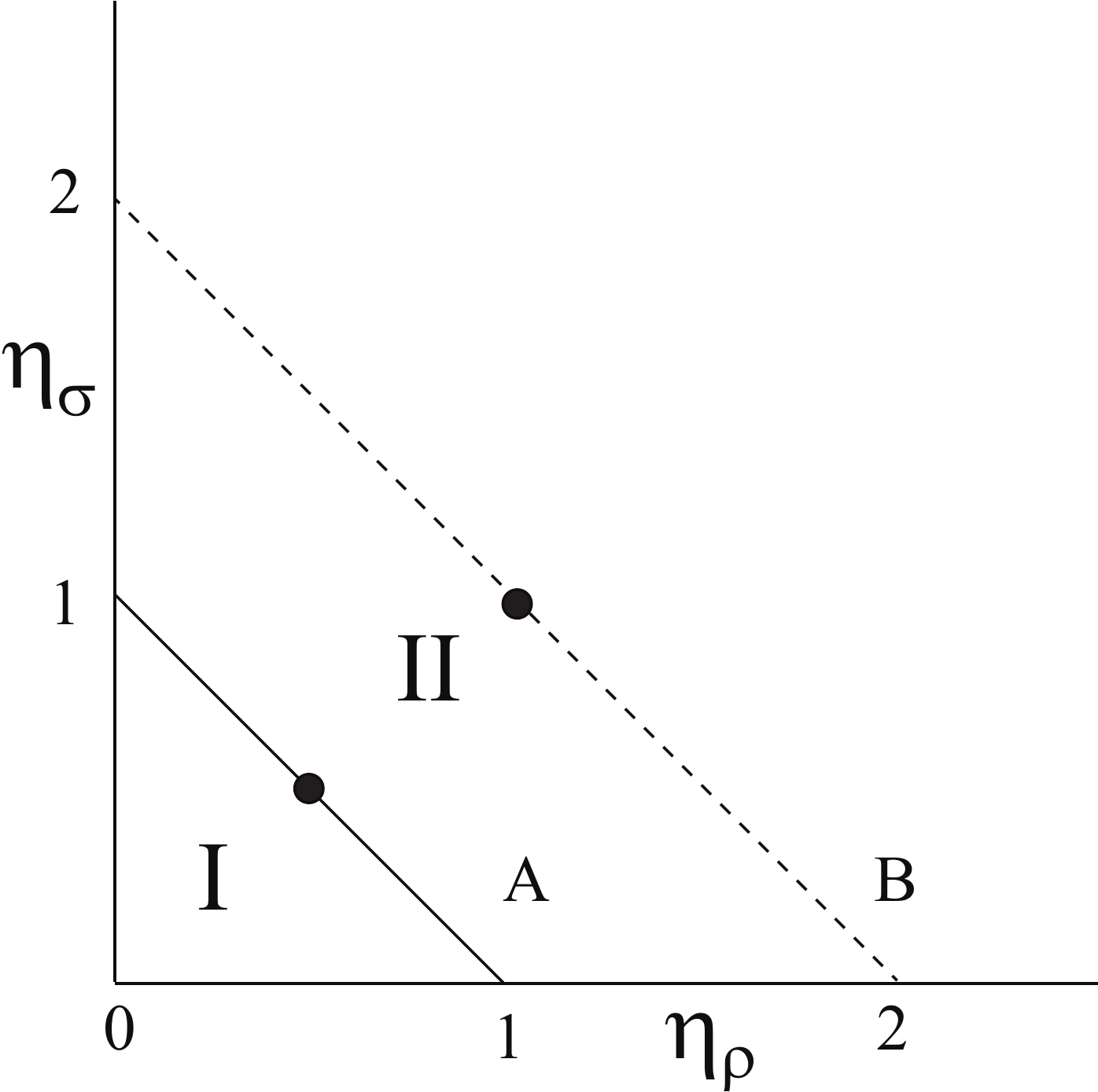}
   \end {center}
   \caption {\label{phase} The phase diagram and the critical Line, $\eta_\rho+\eta_\s =1$.}
\end{figure}

\section{Discussions and Conclusions}

The critical point $\a_\rho = \a_\s = 2$ (equivalently $\eta_\rho = \eta_\s = 1/2$), where the action for the model becomes that of two copies of the critical Schmid models, is found to be on the critical line (critical line A), 
\beq \label{critical}
\frac{1}{\a_\rho} + \frac{1}{\a_\s} = 1, ~~{\rm equivalently}~~ \eta_\rho+ \eta_\s = 1, 
\eeq
as expected. However, it is not on the critical line $\eta_\rho + \eta_\s = 2$ which asserted in the 
previous work. The critical line A Eq.(\ref{critical})  (phase boundary) differs from the critical line
(critical line B)
discussed in refs.\cite{kane:1992a,furusaki}
\beq \label{critical2}
\frac{1}{\a_\rho} + \frac{1}{\a_\s} = 2, ~~{\rm equivalently}~~ \eta_\rho+ \eta_\s = 2.
\eeq

Let us choose a point $\a_\rho = \a_\s = 1$ ($\eta_\rho = \eta_\s = 1$) on the 
critical line B and examine the criticality of this point. 
At this point the action for the model is given as 
\beq
S &=& \frac{1}{4\pi} \int d\t d\s \left[(\p_\t \th)^2 + (\p_\s \th)^2 \right]
+ \frac{1}{4\pi} \int d\t d\s \left[(\p_\t \phi)^2 + (\p_\s \phi)^2 \right] \nn\\
&& +2V_0 \int \frac{d\s}{2\pi} \cos\th \cos\phi \Bigl\vert_{\t=0} .
\eeq
In terms of the two boson fields $\phi_1 = \th + \phi$, $\phi_2 = \th -\phi$, the action can be written as
\beq
S = \frac{1}{4\pi} \int d\t d\s \half \sum_{a=1}^2 \p \phi_a \p \phi_a + V_0 \int \frac{d\s}{2\pi}
\sum_{a=1}^2 \left(e^{i\phi_a} + e^{-i\phi_a}\right)\Bigl\vert_{\t=0}.
\eeq
If we scale the boson fields $\phi_a$ as $\phi_a \rightarrow \sqrt{2} \phi_a$, $a=1,2$, 
\beq
S = \frac{1}{4\pi} \int d\t d\s \sum_{a=1}^2 \p \phi_a \p \phi_a + V_0 \int \frac{d\s}{2\pi}
\sum_{a=1}^2 \left(e^{i\sqrt{2}\phi_a} + e^{-i\sqrt{2}\phi_a}\right)\Bigl\vert_{\t=0}.
\eeq
By applying the Fermi-Bose equivalence, we may attempt to map the action of two boson fields $\phi_a$, $a=1,2$ 
onto a Dirac fermion action of two flavors. But the boundary operator $:e^{\pm i\sqrt{2} \phi_a}:$ has a wrong dimension to be on the boundary. Although in the bulk it has a conformal dimension $(1/2,1/2)$ so that it may be written as a fermion mass operator $\barpsi^a\psi^a$, it has a dimension $(1,1)$ 
thanks to the Neumann boundary condition, which doubles its dimension on the boundary. Hence, its 
dimension does not match that of the fermion bilinear. 
(See more detailed discussion on this point refs.\cite{guinea,Hassel,Polchinski:1994my}.) 
We may rephrase this point in a slightly different way. We may write the operator $:e^{ i\sqrt{2} \phi_a}:
= :e^{ i\sqrt{2} \phi_{aL}}::e^{ i\sqrt{2} \phi_{aR}}:$
as a fermion bilinear $\psi^{a\dag}_L\psi^a_R$, but on the boundary with help of the Neumann boundary condition, it may be also written as 
$-i\psi^{a\dag}_L \psi^{a\dag}_L$, which is a null operator. 
That is, we cannot put the operator $:e^{ i\sqrt{2} \phi_a}:$ as fermion bilinear field operators in a consistent way on the boundary where the Neumann condition is imposed. Thus, the point $\a_\rho=\a_\s =1$ 
($\eta_\rho=\eta_\s =1$) on the line B cannot be the critical point.  


We should also note that it is easy to miss the additional contribution to the kinetic 
action of the boson fields $\th_{ai}$ from the $U_A(1)$ chiral anomaly in the framework of 
the boson theory. It appears only through the non-trivial transformation of the path integral 
measure of the fermi fields under the $U_A(1)$ chiral phase transformation. Let us suppose that 
the additional contribution to the action of $\th_{ia}$ from the chiral anomaly is missed. From Eq.(\ref{bulkspin},\ref{similarity}), we see it is equivalent to replacing $\k^2_1$ and $\k^2_2$ as
$\k^2_1  \rightarrow \frac{\a_\rho -2}{4}$, $\k^2_2  \rightarrow \frac{\a_\s -2}{4}$: 
Then the RG equation would be read as 
\beq
\bolG^{(0)}_{\a\b} [n]+ \bolG^{(1)}_{\a\b} [n] = \frac{V^2_0}{16\pi^2} \Bigl(1 + \half \left(
\frac{\a_\rho}{4} + \frac{\a_\s}{4} -1 \right) \zeta(1)+\cdots \Bigr) G_{\a\b}[n],
\eeq
and the critical line would be obtained as
\beq
\a_\rho + \a_\s = 4,
\eeq
which is certainly erroneous. Thus, the $U(1)$ chiral anomaly plays an important role to fix the critical 
lines of the model. 

We conclude this paper with a remark on the renormalization of composite operators. 
Although the composite operators like four fermi boundary interaction terms,  
$\barpsi_a \psi_a \barpsi_b \psi_b$, $a = 1, 2$ 
cannot be included as a part of the bare boundary action, they may be generated as descendant operators in the 
high order radiative corrections. The renormalizability of the model at higher order may set more constraints 
on the parameters of the model, $\a_\rho$ and $\a_\s$, bringing us richer structure of the phase diagram. 
The full structure of the phase diagram would be accomplished only by examining the renormalization of a complete set of the composite operators \cite{Joglekar}
at a given order. We will discuss renormalilzation of the composite 
operators along the direction of extension of this work elsewhere.

\vskip 1cm

\noindent{\bf Acknowledgments}

This work is supported by Kangwon National University Research Grant 2013. This work is also 
supported in part by the Basic Science Research Institute, Kangwon National University.





\end{document}